\documentclass{elsart}
\usepackage{graphicx}
\usepackage{amssymb}
\usepackage{amsmath}
\usepackage{t1enc} 
\usepackage[utf8]{inputenc} 

\usepackage[]{natbib}
\usepackage{setspace}

\newcommand{\be}{\begin{equation}}
\newcommand{\ee}{\end{equation}}
\newcommand{\ba}{\begin{array}}
\newcommand{\ea}{\end{array}}

\newcounter{ale}

\newenvironment{liste}{\begin{itemize}}{\end{itemize}}
\newcommand{\aliste}{\begin{liste} \setcounter{ale}{1}}
\newcommand{\zliste}{\end{liste}}

\begin{document}

\begin{frontmatter}



\title{Extrinsic noise passing through a Michaelis-Menten reaction: A universal response of a genetic switch}



\author{Anna~Ochab-Marcinek \corauthref{cor1}}
\ead{ochab@ifka.ichf.edu.pl}
\corauth[cor1]{}
\address{Department of Soft Condensed Matter, Institute of Physical Chemistry, 
Polish Academy of Sciences, Warsaw, Poland}

\begin{abstract}

The study of biochemical pathways usually focuses on a small section of a protein interactions network. Two distinct sources contribute to the noise in such a system: intrinsic noise, inherent in the studied reactions, and extrinsic noise generated in other parts of the network or in the environment. We study the effect of extrinsic noise entering the system through a nonlinear uptake reaction which acts as a nonlinear filter. Varying input noise intensity varies the mean of the noise after the passage through the filter, which changes the stability properties of the system.  The steady-state displacement due to small noise is independent on the kinetics of the system but it only depends on the nonlinearity of the input function. 

For monotonically increasing and concave input functions such as the Michaelis-Menten uptake rate, we give a simple argument based on the small-noise expansion, which enables qualitative predictions of the steady-state displacement only by inspection of experimental data: when weak and rapid noise enters the system through a Michaelis-Menten reaction,  then the graph of the system's steady states vs. the mean of the input signal always shifts to the right as noise intensity increases. 

We test the predictions on two models of \textit{lac} operon, where TMG/lactose uptake is driven by a Michaelis-Menten enzymatic process. We show that as a consequence of the steady state displacement due to fluctuations in extracellular TMG/lactose concentration the \textit{lac} switch responds in an asymmetric manner: as noise intensity increases, switching off lactose metabolism becomes easier and switching it on becomes more difficult.

\end{abstract}

\end{frontmatter}

\clearpage
\begin{figure}[t]
\begin{center}
\includegraphics[width=14cm]{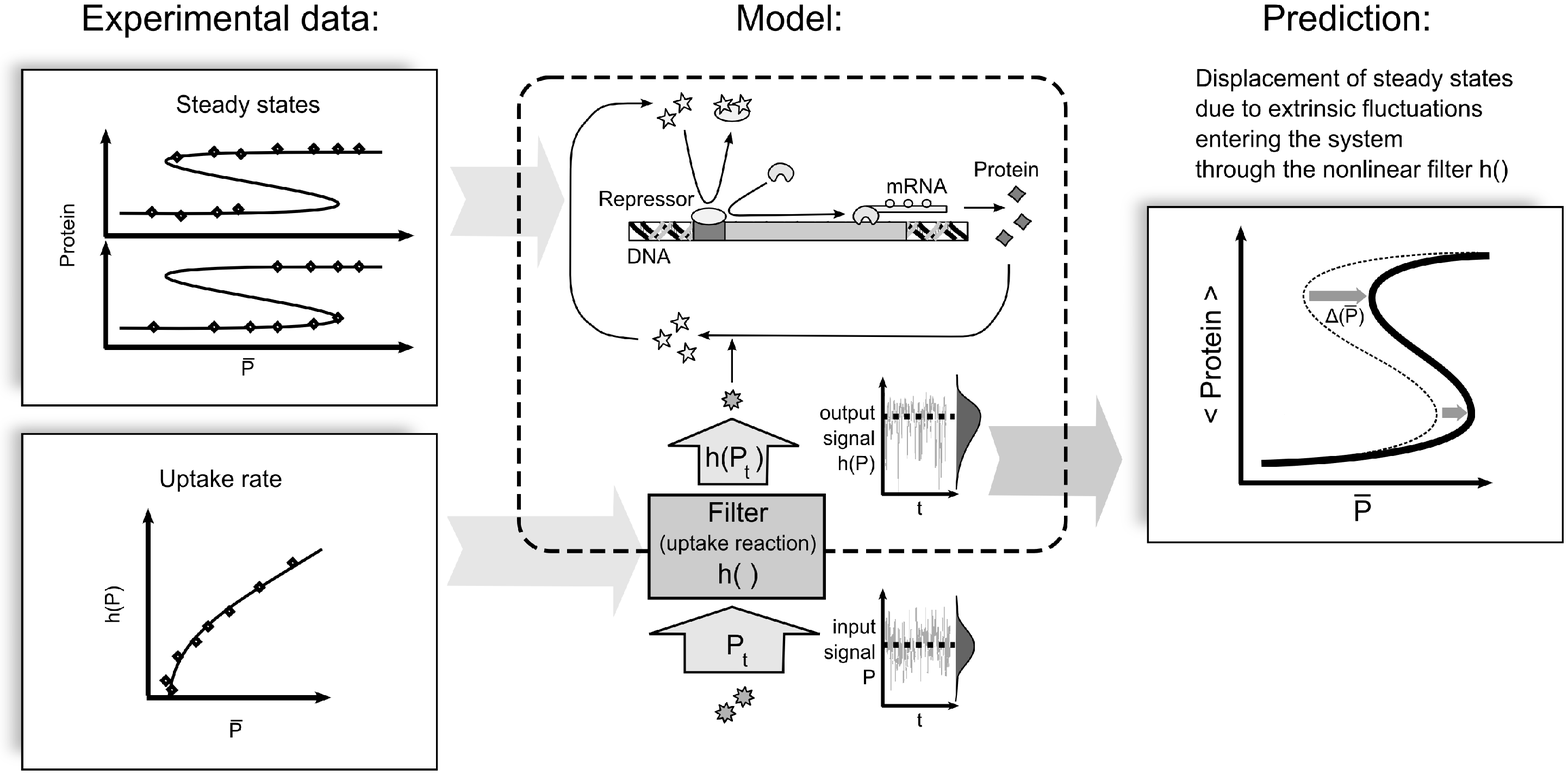}                                     \end{center}
\caption{\label{fig:scheme_noise} Steady-state displacement due to noise passing through a nonlinear uptake reaction. Extrinsic noise $P_t$ enters the system through the uptake reaction which acts as a nonlinear filter $h()$: for example, when the input signal has a Gaussian distribution, the output signal would have a skewed distribution. For small noise, varying the intensity of the input noise, without changing its mean $\bar{P}$, varies the mean of the output noise $\langle h(P_t) \rangle$. This causes a shift $\Delta(\bar{P})$ of the steady states of the system. The displacement direction depends on the shape of the filtering function $h(P_t)$. The small-noise expansion method enables qualitative prediction of the displacement direction, by inspection of the following experimental data (here, an example sketch): the graph of the system's steady states vs. $\bar{P}$  and the graph of the uptake rate vs. $\bar{P}$. Real experimental data can be found e.g. in \citep{Ozbudak}.}
\end{figure}

\section{Introduction}
When studying biochemical pathways, we usually focus on a very small section extracted from a large network of protein interactions.
Two distinct sources contribute to the noise in the studied system: Intrinsic noise due to the randomness of molecular collisions which result in chemical reactions of interest, and extrinsic noise, generated in other parts of the network or in the external environment \citep{Shibata}. 

Computational methods developed based on the Gillespie algorithm \citep{Gillespie1976, Gillespie2007} enormously increased the popularity of theoretical studies of intrinsic stochasticity in biochemical reactions (see the most cited papers: \cite{Thattai2001, Rao2002, OzbudakNatGen2002}). On the other hand, growing attention is being paid to the study of extrinsic noise  \citep{Swain_Elowitz,Raj,Shahrezaei_Swain,Shahrezaei_Ollivier, Lei}, in particular to the problem of discrimination between the effects of intrinsic and extrinsic noise components in studied systems  \citep{Elowitz,Raser,Paulsson2004,Pedraza2005,Newman2006}. Experiments on \textit{lac} gene in \textit{E. coli} \citep{Elowitz} and \textit{PHO5} and \textit{GAL1} genes in \textit{Saccharomyces cerevisiae} \citep{Raser} have demonstrated that the contribution of extrinsic noise to gene expression may be significant.

Noise propagates across a gene network by passing from one subsystem to the other through reactions which link particular sections of the network, or which link the network with the external environment. In this paper, we study the effect of extrinsic fluctuations entering the studied system through a nonlinear uptake reaction which acts as a nonlinear filter. In this way, one of the reaction rates in the system depends in a nonlinear way on a fluctuating extrinsic variable.  \cite{Shahrezaei_Ollivier} simulated (with a Gillespie algorithm modified by inclusion of a randomly varying reaction rate) a simple general model of gene expression where one of the reaction rates was an exponential function of a slowly fluctuating Gaussian process. The resulting reaction rate had an asymmetric, log-normal distribution. This particular shape of the nonlinear filter was chosen to conform the experimentally measured distributions   of gene expression rates \citep{Rosenfeld}. \cite{Shahrezaei_Ollivier} found that the presence of this kind of noise shifts the mean concentration of the reaction product, however they did not explain that effect with analytical calculations.

While \cite{Shahrezaei_Ollivier} studied numerically the effect of extrinsic fluctuations whose average lifetime was longer than the shortest timescale of the system (thus interfering the relaxation dynamics of the system's variables), we focus on the analytical study of the effect of weak and rapid extrinsic noise, whose instantaneous fluctuations are too fast to directly interfere the system's dynamics. The assumption of weakness and rapidity of the noise allows to use the small-noise expansion method \citep{Gardiner}. A measure for the rapidity of the random fluctuations is the correlation time \citep{Horsthemke}. When the characteristic time scale of the noise (the correlation time) is much faster than the time scale of the reactions within the system of interest, then the noisy input signal contributes to the kinetics of the system as an average signal only \citep{Horsthemke}. One could expect that, in such a case, varying the input noise intensity  does not influence the behavior of the system. However, the non-linear filtering function can transform the output noise distribution in such a way that its mean varies as the input noise intensity is varied (see Fig. \ref{fig:scheme_noise}). The system may therefore respond to extrinsic noise by shifting its steady states with respect to the deterministic ones \citep{Ochab,Rocco,Gerstung}. 

A particularly interesting case is a multistable system, for example a bistable switch \citep{Laurent_Kellershohn,Ferrell,Wilhelm}, a chemical reaction system where two distinct steady states are possible under the same external conditions. In the presence of a noisy input signal, the bistable region can move to a concentration range where otherwise only one steady state was present, or bistability can disappear for concentrations at which the system was initially bistable \citep{Ochab}. Thus, as noise intensity increases, the existing steady states of the bistable system can be stabilized or destabilized, which  results in the increase or decrease of the escape time from one steady state to the other \citep{Ochab}. The switching time in a bistable genetic switch has been recently investigated by \cite{Cheng}, as the measure for the robustness of the cellular memory to intrinsic noise. The simplicity of the system (one-dimensional equation of kinetics) allowed them for analytical calculation of the switching time as the mean first-passage time derived from the Fokker-Planck equation \citep{Gardiner}. At a fixed noise level,  \cite{Cheng} studied the dependence of the switching time on the transcription rate. On the other hand, \cite{Lei} simulated a similar system with a  transcription delay. Extrinsic noise was there applied in the same log-normal form as in the above-mentioned paper by \cite{Shahrezaei_Ollivier} and the noise was strong enough to allow for bi-directional switching.  \cite{Lei} showed that the switching frequency decreases as the delay in the negative feedback loop increases.

The study of stochastic control of metabolic pathways from the perspective of the small-noise expansion is relatively novel. It was initially presented in \citep{Ochab} where we used the small-noise expansion to calculate the noise-induced steady-state displacement for an arbitrary filtering function and tested it on the reduced Yildirim-Mackey model  \citep{Yildirim_Mackey} of the \textit{lac} operon. Using a numerical simulation of that model, we showed that the induction time of the operon increases and the uninduction time decreases as a consequence of the noise-induced displacement. Recently,  \citep{Rocco} studied fluctuations in enzyme activity in Michaelis-Menten kinetics, expressing the noise-induced steady-state displacement as the Stratonovich drift. On the other hand, \citep{Gerstung} used the small-noise expansion method to study the steady-state displacement of concentrations of cis-regulatory promoters due to intrinsic fluctuations in transcription factor concentration combined with extrinsic fluctuations in the transcription factor synthesis rate. The non-linear input function (the filter) was the probability of promoter occupation in the form of $h(n) = n/(K+n)$, as a function of the fluctuating number $n$ of free transcription factors. 

While in \citep{Gerstung} the above form of the filtering function arised from the stationary solution of Master equation for the probability of finding $n$ free transcription factors, in my present work the filtering function of the same form is the uptake rate of  a certain substance P (the input signal), which enters the system through a catalytic Michaelis-Menten reaction (see Fig. \ref{fig:scheme_noise}). In a reaction that follows  Michaelis-Menten kinetics, the reactant combines reversibly with an enzyme to form a complex. The complex then dissociates into the product   and the free enzyme. If the total enzyme concentration does not change over time, and the concentration of the substrate-bound enzyme changes much more slowly than those of the product and substrate (the quasi-steady-state assumption), then the intermediate enzymatic steps can be reduced in the equations of chemical kinetics. The rate of product formation (P uptake) is then proportional to $h(P) = P/(K_M+P)$, where $P$ is the concentration of P  \citep{Atkins}.

We show that, when the P concentration $P_t$ fluctuates with a Gaussian distribution, then $h(P_t)$ has an asymmetric distribution, similarly as in \cite{Shahrezaei_Ollivier} and \cite{Lei}. A simple calculus argument, based on the small-noise expansion method, allows to show that the noise-induced steady-state displacement depends solely on the shape of the nonlinear input function \citep{Gerstung}. Consequently, we argue that the Michaelis-Menten type uptake function (monotonically increasing and concave) always generates the same type of the steady-state displacement. We show that this argument enables qualitative predictions of the steady-state displacement only by inspection of experimental data, i.e. a) the graph of  the uptake function vs. the mean of the input signal and b) the graph of system's steady states vs. the mean of the input signal. 

 We show how, as a consequence of the steady-state displacement, the induction/uninduction time of the \textit{lac} operon changes depending on the intensity of TMG/lactose fluctuations. To illustrate the universality of the phenomenon, we compare the results for the Ozbudak model of \textit{lac} operon \citep{Ozbudak}, where the extracellular TMG uptake rate is an experimentally fitted function $\sim P^{0.6}$ (increasing and concave), with the results for the reduced Yildirim-Mackey model \citep{Ochab, Yildirim_Mackey}, where the extracellular lactose uptake rate is given by the Michaelis-Menten formula $P/(K_M+P)$.  For both models we show how the Gaussian distribution of the input signal generates the skewed distribution of the Michaelis-Menten uptake rate. We analyze the validity range of the small-noise expansion in both cases. We compare the noise-induced steady state displacement and the noise-induced changes in the induction/uniduction time for the Ozbudak model with the  previously published results for the reduced Yildirim-Mackey model \citep{Ochab}.

\section{Theory}

\subsection{Small-noise expansion method}
The input concentration of a substance P is modeled by the stochastic process $P_t$ with the mean $\bar{P}$ and the variance $\sigma^2$. The passage through the uptake reaction generates the output noise $h(P_t)$.

The equations of kinetics of the studied system are:
\be \label{eq:stoch_system}
\dot{\vec{X}} = \vec{F}(\vec{X}, h(P_t)),
\ee
where $\vec{X}$ is the vector of concentrations of substrates, and the system depends on $P_t$ through the function $h(P_t)$ only.

Assume that the noise is weak and slower than the time scale of the filter response but faster than the characteristic time scale of the system (\ref{eq:stoch_system}). Then the system experiences the output noise (after the passage through the filter) as its mean, $\langle h(P_t) \rangle$ \citep{Horsthemke}. If the mean value of the output noise differs from the deterministic value of $h(\bar{P})$, then the steady states of
\be
\dot{\vec{X}} = \vec{F}(\vec{X},\langle h(P_t) \rangle)
\ee
are shifted with respect to steady states of the deterministic system 
\be
\dot{\vec{X}} = \vec{F}(\vec{X}, h(\bar{P})).
\ee
It is possible to approximately find the displacement without the knowledge of the equations of kinetics, just by transformation of the $X$ vs. $\bar{P}$ graph. While the deterministic system has stationary states $\vec{X}^*$, the stochastic system has quasi-steady states $\langle \vec{X}\rangle$, around which its trajectories fluctuate (assuming that a noise-induced transition between multiple steady states is very unlikely). In the stochastic system, the steady states differ from the deterministic ones. This difference can be graphically shown as the shift of the graph of the steady states $\vec{X}^*$ vs. $\bar{P}$ along the $\bar{P}$ axis \footnote{There are two reasons to consider the horizontal shift (with respect to  $\bar{P}$) and not the vertical one (with respect to $\langle \vec{X} \rangle$): a) The input signal $P_t$ is the primary source of noise so it is easier to perform the calculations based on the dependence of the steady states on the variations in $P_t$ than on the variations of $\vec{X}$, which anyway depend those of $P_t$. b) We will analyze the noise-induced shift in bistable systems, which can have two steady states for one value of $\bar{P}$, so the vertical shift would be ambiguous.} (see Fig. \ref{fig:scheme_noise})  by a function $\Delta(\bar{P})$, such that
\be
\langle \vec{X}(P_t) \rangle = \vec{X}^*(\bar{P} + \Delta(\bar{P})).
\ee
The system depends on $P_t$ only in the function of the output process $h(P_t)$, so
\be \label{eq:xh}
\langle \ \vec{X} (h(P_t)) \ \rangle = \vec{X}^*( \ h(\bar{P} + \Delta(\bar{P})) \ ).
\ee
As assumed above, the system only experiences the mean of the output process (Fig. \ref{fig:scheme_noise}):
\be \label{eq:xrang}
\langle \ \vec{X} (h(P_t)) \  \rangle = \vec{X}^*( \ \langle h(P_t) \rangle \ ),
\ee
And therefore, from the combination of Eqs. (\ref{eq:xh}) and (\ref{eq:xrang}):
\be
 \langle h(P_t) \rangle = h(\bar{P} + \Delta(\bar{P})).
\ee
In other words, when the input process $P_t$ is stochastic (fluctuating around the mean $\bar{P}$), then the mean of the output process $h(P_t)$ is equal to the deterministic output under the shift $\Delta(\bar{P})$ of the input flux.

Assume that $\Delta(\bar{P})$ depends on $\bar{P}$ so weakly that it can be treated as a constant $\Delta$.  
Expand the output process $h(\bar{P} + \Delta)$ around the mean of the input process:
\be \label{eq:expansion_Delta}
 \langle h(P_t) \rangle \approx  h(\bar{P} + \Delta) = h(\bar{P}) + h'(\bar{P}) \Delta + ...
\ee
(As shown in the Results section and in \citep{Ochab}, this approximation works well for the example systems under study. In case when $\Delta(\bar{P})$ depends steeply on $\bar{P}$, the approximation may break down.)

On the other hand, expansion of $\langle h(P_t) \rangle $ around $\bar{P}$ with respect to a fluctuation $\delta P$ of the input process ($P_t=\bar{P}+\delta P$ at a given time $t$)  yields: 
\be \label{eq:expansion_fluct}
 \langle h(P_t) \rangle=\langle h(\bar{P}+\delta P) \rangle = h(\bar P) + h'(\bar P)\langle \delta P \rangle + \frac{1}{2}h''(\bar P)\langle \delta P^2 \rangle + ...
\ee
But the mean deviation from the mean of the input process $\langle \delta P \rangle = 0$ and the variance of the input process $\langle \delta P^2 \rangle=\sigma^2$. Combination of Eqs. (\ref{eq:expansion_Delta}) and (\ref{eq:expansion_fluct}) yields  the approximate formula for the noise-induced shift of steady states with respect to deterministic steady states:
\be \label{eq:Delta}
\Delta(\bar{P}) = \frac{h''(\bar{P})}{2 h'(\bar{P})} \sigma^2
\ee

\subsection{Michaelis-Menten uptake rate as the non-linear filter}

In the case of a system where the filtering function is generated by Michaelis-Menten kinetics \citep{Atkins}, the small-noise expansion method is valid when the noise is slower than the intermediate enzymatic reactions, which had been reduced into $h(\bar{P})$. Then, the enzymatic reactions are not perturbed by the fluctuations in P concentration, and the concentrations of the reactants can reach their steady state while the P concentration is approximately constant.

The noise correction $\Delta (\bar{P})$ to the Michaelis-Menten rate $h(\bar{P}) = \bar{P}/(K_M+\bar{P})$ can be viewed as a correction to the Michaelis constant $K_M$. The displacement $\bar{P}+\Delta(\bar{P}) = \bar{P} (1 + \Delta(\bar{P})/\bar{P}) $ is equivalent to the correction
\be \label{eq:Knoise}
K_{M,noise} (\bar{P})=  \frac{K_M}{ 1+\frac{\Delta(\bar{P})}{\bar{P}}} = \frac{K_M}{1 - \frac{\sigma^2}{(K+\bar{P})\bar{P}}}.
\ee 

\begin{figure}[t]
\begin{center}
\includegraphics[width=6cm]{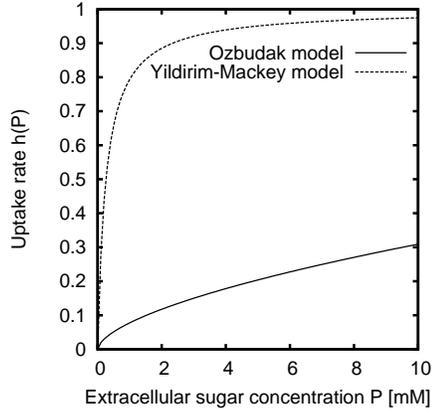}
\end{center}
\caption{\label{fig:uptake} The TMG uptake rate (\ref{eq:tmg_uptake}) for the Ozbudak model (solid line) and the lactose uptake rate (\ref{eq:lactose_uptake}) for the Yildirim-Mackey model (dashed line). The mathematical formulas for the uptake rates are different in both models, but as they describe the same uptake mechanism, their graphs have a similar Michaelis-Menten shape with $h'>0$ and $h''<0$, which generates the same type of the steady-state shift due to noise in extracellular TMG/lactose concentration (see the Results section).}
\end{figure}

\begin{figure}[t]
\begin{center}
\includegraphics[width=6.7cm]{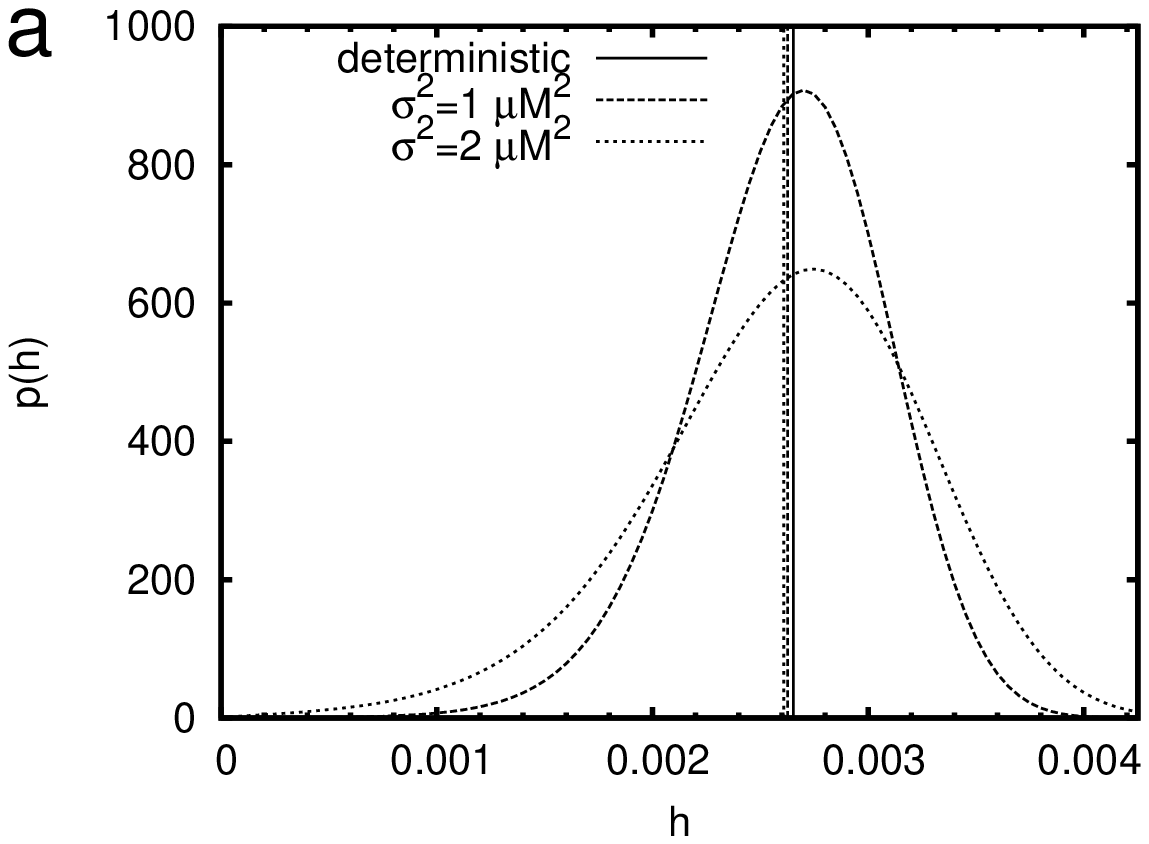} $ \ \ $ \includegraphics[width=6.7cm]{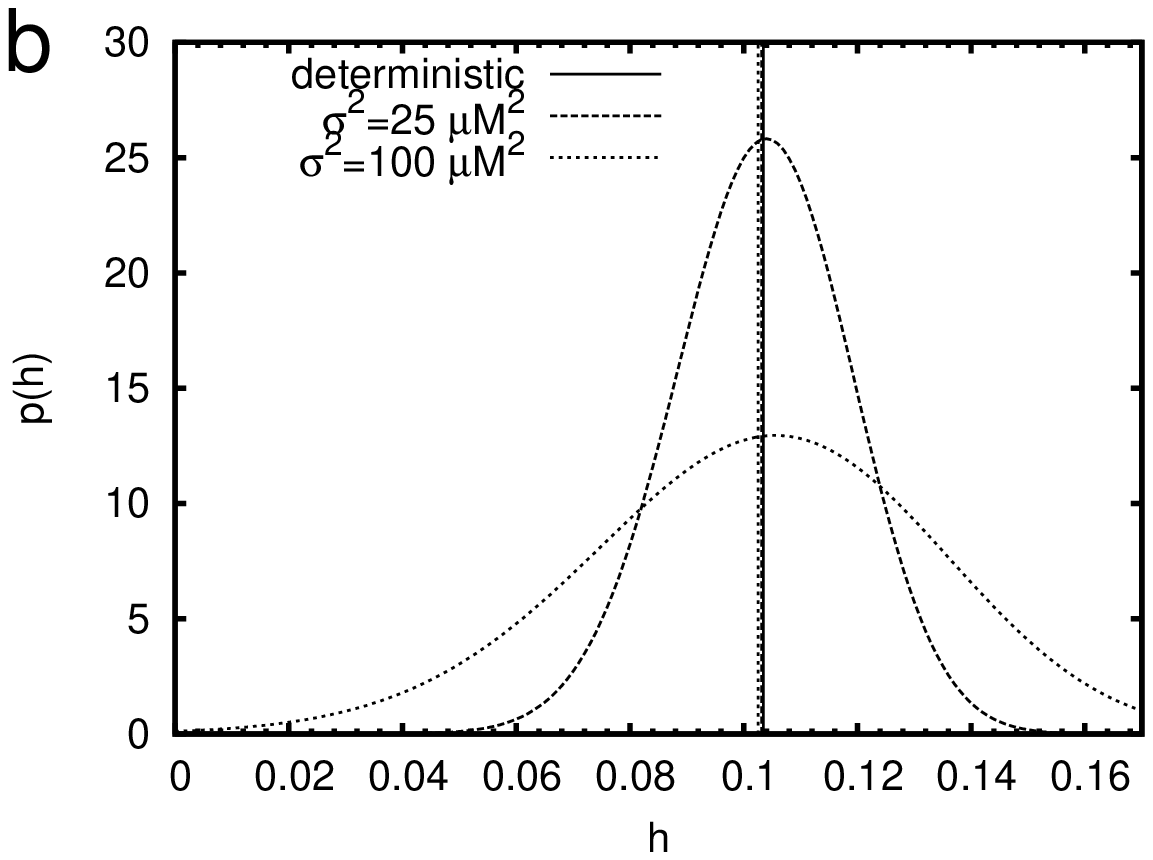}
\end{center}
\caption{\label{fig:prob} Probability density functions for the TMG/lactose uptake rates when the fluctuating concentrations of the extracellular TMG/lactose have a Gaussian distribution with the variance $\sigma^2$. Vertical lines indicate the mean, which is slightly shifted with respect to the deterministic uptake rate (solid line). a) Ozbudak model. b) Yildirim-Mackey model.}
\end{figure}

When the input signal $P_t$ has a Gaussian distribution with the mean $\bar{P}$ and variance $\sigma^2$, the output signal $h(P_t) = P_t/(K+P_t)$ (Fig. \ref{fig:uptake}) has an asymmetric probability density function (Fig. \ref{fig:prob} b):
\be \label{eq:ph}
p(h) = \frac{K}{\sqrt{2 \pi \sigma^2} (1-h)^2} \exp\left[-\frac{1}{2 \sigma^2}\left( \frac{K h}{1-h}-\bar{P} \right)^2\right]
\ee
(See the Appendix \ref{app:pdf} for the details of the calculation.)
The mean $\langle h(P_t) \rangle = \int  h p(h) dh /  \mathcal{N}$ of (\ref{eq:ph}) exists if $0\geq h \geq 1$, which corresponds to $0 \geq \bar{P} \geq +\infty$. The normalization constant $\mathcal{N} = (1/2) (\mathrm{erf}[\bar{P} \sqrt{2 }  / (2 \sigma)]+1)$, and the accurate values of $\langle h(P_t) \rangle$ can be computed numerically.

\section{Models}
 The \textit{lac} operon is one of the most extensively studied examples of a bistable genetic switch. In a certain range of extracellular TMG/lactose concentration, the gene transcription can be in one of two discrete states, either fully induced or uninduced. Bistability is generated by the positive feedback loop, where high lactose/TMG concentration in the cell causes weak repression of the \textit{lac} gene transcription and, in consequence, more permease is produced which pumps more lactose/TMG into the cell.  According to experimental data  \citep{Huber, Wright, Page, Lolkema, Ozbudak} the rate of the TMG/lactose uptake into the cell is of Michaelis-Menten type. Extracellular lactose concentration experienced by \textit{E. coli} living in its natural environment may fluctuate, due to high mobility of the bacterium (see e.g. \cite{DiLuzio}), granularity of the intestinal content and motions of intestinal villi.

\subsection{Ozbudak model}\label{subsec:ozbudak}

The model \citep{Ozbudak} is based on the experiment where a gene coding for a fluorescent protein was incorporated under the control of the \textit{lac} promoter in E. coli bacteria, in order to indicate the transcription activity of the  \textit{lac} gene. The equations of chemical kinetics for this system are the following:
\be \label{eq:ozbudak}
\frac{R}{R_T}=\frac{1}{1+(x/x_0)^2}
\ee
\be
\tau_y \frac{dy}{dt}=\alpha \frac{1}{1+R/R_0}-y
\ee
\be
\tau_x \frac{dx}{dt}= \beta_G h(T) y - x
\ee
$x$ denotes the intracellular TMG concentration, $y$ is the
concentration of permease in green fluorescence units, $R_T$ is the total
concentration of the repressor, and $R$ is the concentration of
active repressor. The active fraction of the repressor is a function of the TMG concentration $x$, with half-saturation
concentration $x_0$. $\alpha$ is the maximum rate of generation of permease, $R_0$ is the half-saturation of the repressor. Permease is
depleted in a time scale $\tau_y$, due to a combination of degradation and dilution. TMG enters the cell at the rate $h(T)$ (Fig. \ref{fig:uptake}) proportional to $y$ and to the glucose uptake rate $\beta_G$, and is depleted with time constant $\tau_x$. 

Steady states of $y$ (in $x_0$ units, further on called the 'scaled units') are given by:  
\be\label{eq:ozbudak_steady}
y=\alpha\frac{1+(\beta_G h(T) y)^2}{1 + \frac{R_T}{R_0} + (\beta_G h(T) y)^2}
\ee
$T$ is measured in $\mathrm{\mu M}$. In a certain range of TMG concentrations, the cubic equation (\ref{eq:ozbudak_steady}) has three roots (two stable fixed points separated by one unstable fixed point), which very well reproduces the experimentally observed switch-like behavior of \textit{lac} operon. In this study, we assume that the extracellular glucose concentration is zero. Then, the system is bistable for $3.39<T<24.40 \ \mathrm{\mu M}$. We have chosen the value of $\tau_x=1 \ \mathrm{min}$ significantly larger than the time scale of the noise (\ref{eq:tau_ou}), and at the same time much smaller than $\tau_y=216 \ \mathrm{min}$, which conforms the experimental results \citep{Mettetal} reporting $\tau_x$  less than the measurement resolution of $35 \ \mathrm{min}$.

The TMG uptake rate function
\be \label{eq:tmg_uptake}
h(T) = 1.23 \times 10^{-3} T^{0.6},
\ee
as well as the values of the parameters (Table \ref{tab:ozbudak_params}), have been fitted from the experimental data \citep{Ozbudak}. Since the formula (\ref{eq:tmg_uptake}) is the result of fitting, it does not have the classical Michaelis-Menten form $T/(K+T)$, but its graph has the characteristic Michaelis-Menten shape, increasing and concave.

\subsection{Yildirim-Mackey model}\label{subsubsec:mackey}
The Yildirim-Mackey model \citep{Ochab,Yildirim_Mackey} consists of three equations of chemical kinetics for mRNA ($M$), allolactose ($A$) and lactose ($L$) concentrations in the \textit{E. coli} cell:
\be \label{eq:mysys}
\ba{lll}
\frac{dM}{dt} & = & \alpha_M \ \frac{1+K_1 \ A^2}{1+K_2 R_{tot} + K_1 \ A^2} + \Gamma_0 - \tilde{\gamma}_M \ M \\
&&\\
\frac{dA}{dt} & = & k_B \ M \left(\alpha_A \ \frac{L}{K_L + L} - \beta_A \ \frac{A}{K_A + A}\right) - \tilde{\gamma}_A \ A \\
&&\\
\frac{dL}{dt} & = & k_P \ M \left(\alpha_L \ h(L_e) - \beta_L \ \frac{L}{K_{L1} + L}\right) - \alpha_A \ k_B \ M \ \frac{L}{K_L + L} - \tilde{\gamma}_L \ L \\
\ea
\ee

$\alpha$ and $\beta$ denote the gain and loss rates for the reactions. $K_1$ is the equilibrium constant for the repressor-allolactose reaction. $K_2$ is the equilibrium constant for the operator-repressor reaction, and $R_{tot}$ is the total amount of the repressor. The $\tilde{\gamma} = \gamma + \mu$ are the coefficients for the terms representing decay of species due to chemical degradation ($\gamma$) and dilution ($\mu$). Even if allolactose is totally absent, on occasion repressor will transiently not be bound to the operator and RNA polymerase will initiate transcription. Although the mRNA production rate $dM / dt$ would be then nonzero (a leakage transcription), it is necessary to add an empirical constant $\Gamma_0$ to the model to obtain a leakage rate that agrees with experimental values \citep{Yildirim_Mackey}. The allolactose gain and loss rates and the lactose loss rate depend on  the $\beta$-galactosidase concentration (an anzyme breaking down lactose into allolactose), which is proportional ($k_B$ factor) to the mRNA concentration. Similarly, the lactose gain and loss rates depend on the permease concentration, proportional ($k_P$ factor) to the mRNA concentration. See Table \ref{tab:mackey_params} for the values of parameters.

The system is bistable for $27.7 \ \mathrm{\mu M} \ < L_e < 61.8 \ \mathrm{\mu M}$. The lactose uptake rate function
\be \label{eq:lactose_uptake}
h(L_e) = \frac{L_e}{K_{L_e} + L_e}
\ee
has the Michaelis-Menten form, to conform the data of \cite{Lolkema, Huber, Page, Wright}. 

\begin{figure}[t]
\begin{center}
\includegraphics[width=6cm]{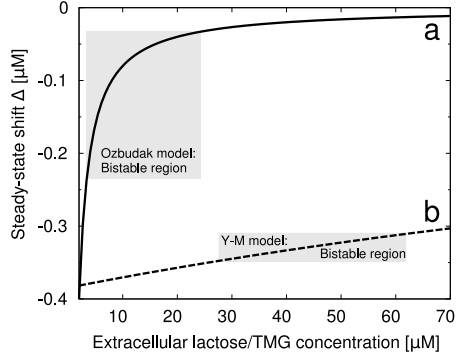}                                     \end{center}
\caption{\label{fig:delta} Comparison of the steady-state displacements calculated using the small-noise expansion method for the Ozbudak model (a) and the Yildirim-Mackey model (b). Rectangles indicate bistable regions of both models. In the Ozbudak model, the noise-induced shift of steady states is weaker than in the Yildirim-Mackey model and it depends more strongly on the extracellular sugar concentration. The displacement is the strongest on the left boundary of the bistable region and the weakest on its right boundary.}
\end{figure}

\subsection{Fluctuations}

The fluctuations in TMG/lactose concentration have been modeled by Ornstein-Uhlenbeck noise. For the small-noise expansion to be valid, the correlation time of the noise 
\be \label{eq:tau_ou}
\tau_{OU}= 1.2 \ \mathrm{s}
\ee  
has been chosen significantly larger than the time scale of the enzymatic TMG/lactose uptake by permease ($0.1 \ \mathrm{s}$, according to \cite{Wright1986}), but smaller than the fastest time scale of the system: $\tau_{\mathrm{sys}} = \tau_x =  1 \ \mathrm{min}$ for the Ozbudak model and $\tau_{\mathrm{sys}} \approx 10^{-1} \ \mathrm{min}$ for Yildirim-Mackey model \citep{Yildirim_Mackey}. Taking into account the bacterial motility (mean velocity of \textit{E. coli} is $\sim 30 \mu \mathrm{m/s}$, \citep{DiLuzio}), the granularity of the intestinal content and motions of intestinal villi, one can suppose that the fluctuation rapidity assumed here is realistic \citep{Ochab}.

The details of the numerical simulation of the stochastic process are presented in the Appendix \ref{app:simul}.

\section{Results}\label{sec:results}
\subsection{Asymmetric distribution of the Michaelis-Menten uptake rate}
\begin{figure}[t]
\includegraphics[width=7cm]{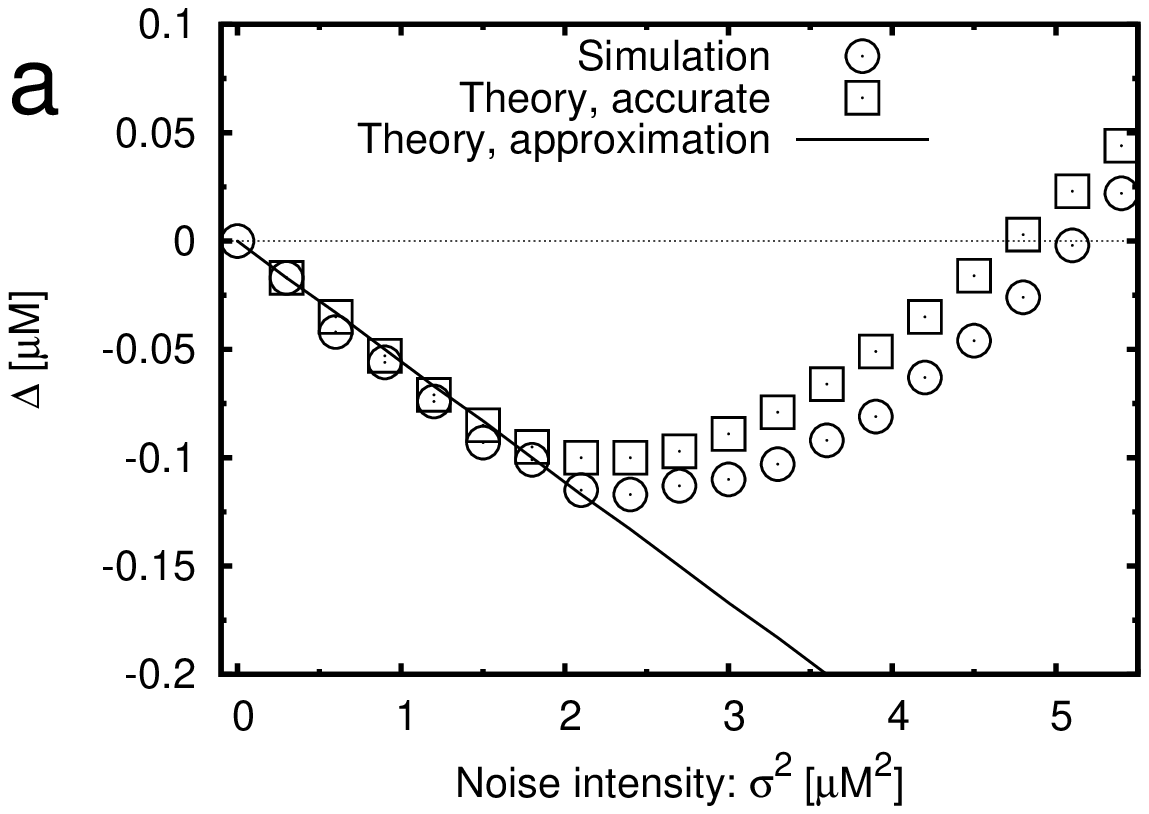} \hspace{0.5cm}
\includegraphics[width=7cm]{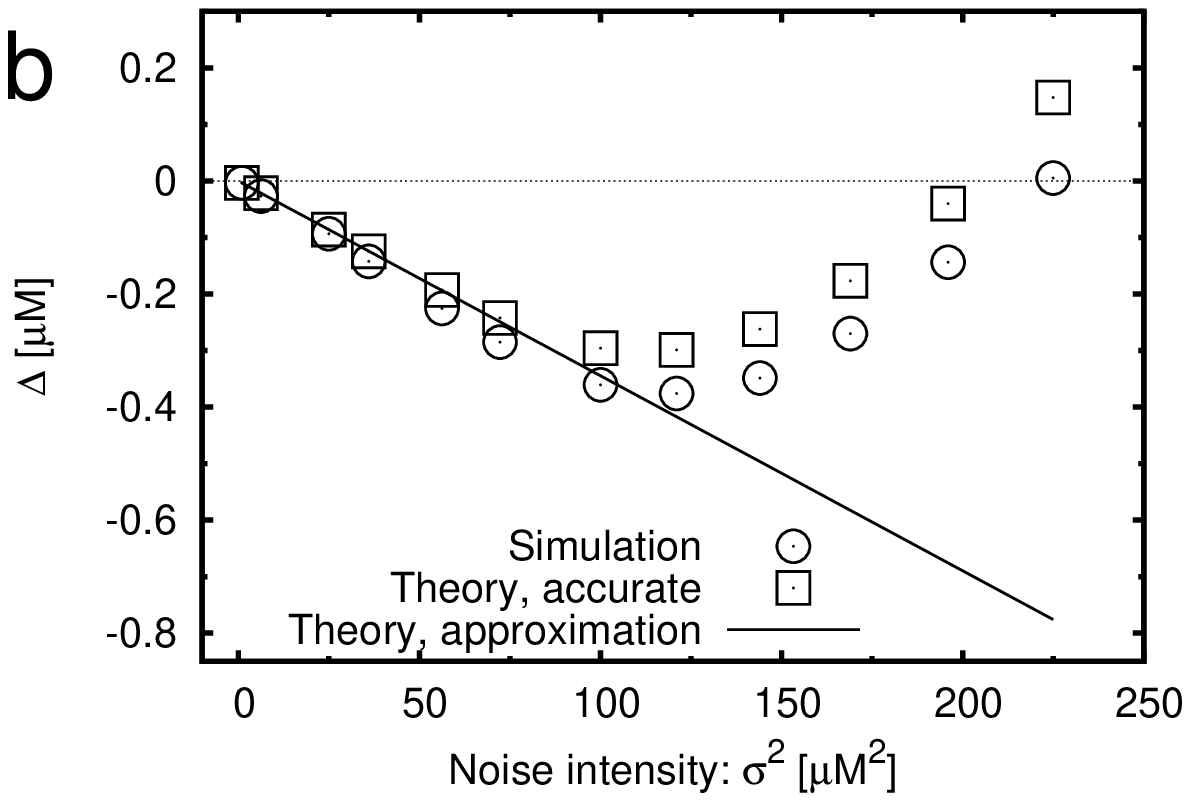}\\
\includegraphics[width=7cm]{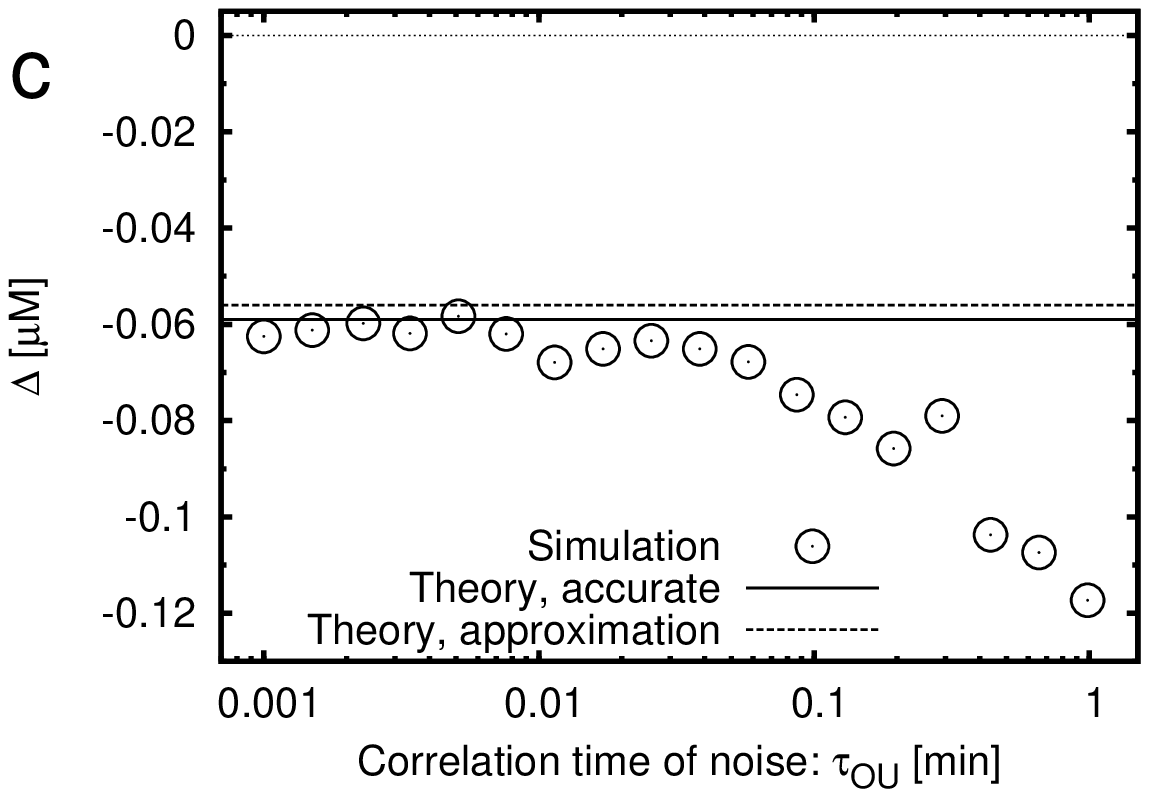} \hspace{0.5cm}
\includegraphics[width=7cm]{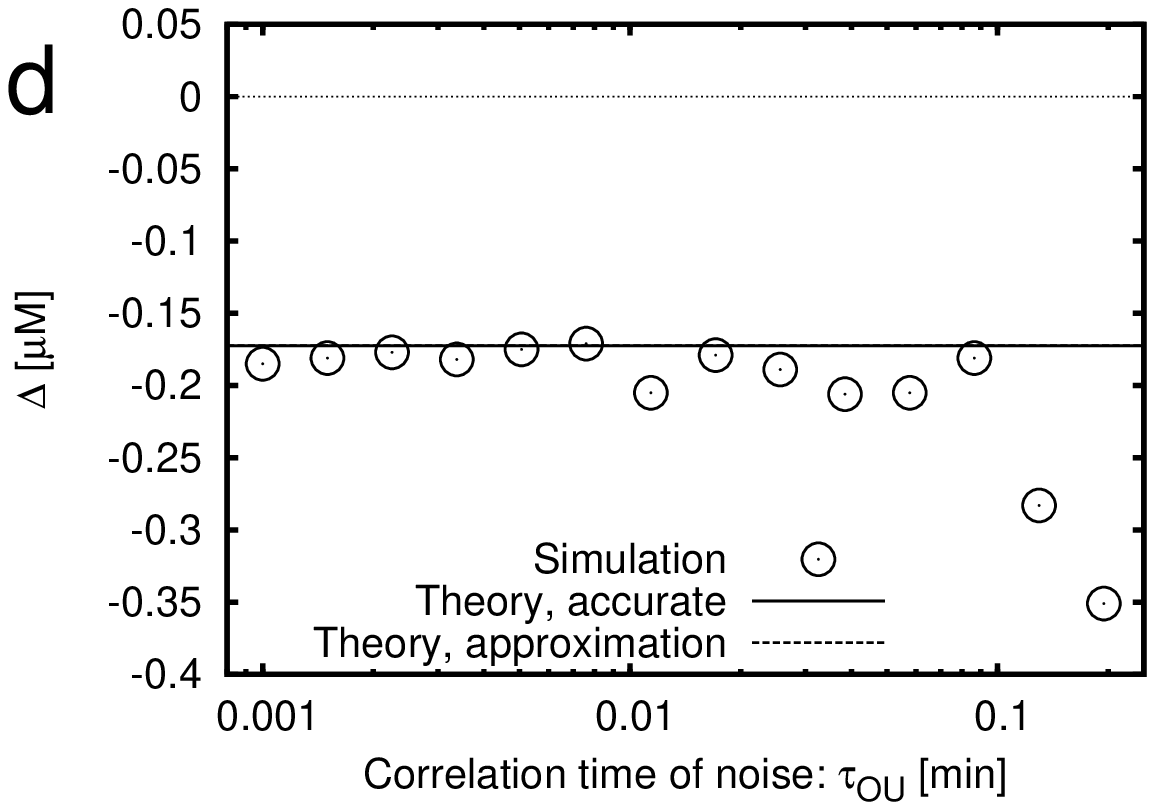}\\
\caption{\label{fig:delta_validity} The range of validity of the small-noise expansion: Comparison of the steady-state displacement calculated using different methods (simulation, accurate calculation, small-noise expansion) for constant values $\bar{T}$ or $\bar{L}_e$ as noise intensity and time scale of the noise are changed. Top panels: Varying noise intensities $\sigma^2$. a) Ozbudak model, $\bar{T}=3.6 \ \mu \mathrm{M}$, $\tau_{OU} = 1.2 \ \mathrm{s} = 0.02 \ \mathrm{min}$. b) Mackey model, $\bar{L}_e = 30 \ \mu \mathrm{M}$, $\tau_{OU} = 1.2 \  \mathrm{s}= 0.02 \  \mathrm{min}$. Bottom panels: Varying noise time scales $\tau_{OU}$. c)  Ozbudak model, $\bar{T}=3.6 \ \mu \mathrm{M}$, $\sigma^2 = 1 \  \mu{M}^2$. d) Mackey model, $\bar{L}_e = 30 \  \mu \mathrm{M}$, $\sigma^2 = 50 \  \mu{M}^2$ (the accurate value overlaps with the approximated one).}
\end{figure}

For the Mackey model, the probability density function of the Michaelis-Menten uptake rate $h$ is given by the formula (\ref{eq:ph}). For the Ozbudak Model, the probability density function of the uptake rate is described by a different formula (see the Appendix \ref{app:pdf} for the details of the calculation):
\be \label{eq:pho}
p(h) = \frac{5}{3} \frac{ h^{2/3}}{b^{5/3} \sqrt{2\pi \sigma^2}} \exp \left[ -\frac{1}{2\sigma^2} \left( \left(  \frac{h}{b}\right)^{\frac{5}{3}} - \bar{T} \right)^2   \right]
\ee
with $b = 1.23\times 10^{-3}$, because $h(T)$ is given by the fitted function (\ref{eq:tmg_uptake}) which, however, has a similar shape to the Michaelis-Menten function. Consequently, the probability density functions for both models also have similar asymmetric shapes, and for small noise their mean values decrease as the noise intensity increases (Fig. \ref{fig:prob}). For larger noise, the mean values begin to increase (compare Fig. \ref{fig:delta_validity}). 

\subsection{Calculation of the steady-state displacement using small-noise expansion}
\begin{figure}[t]
\includegraphics[width=7cm]{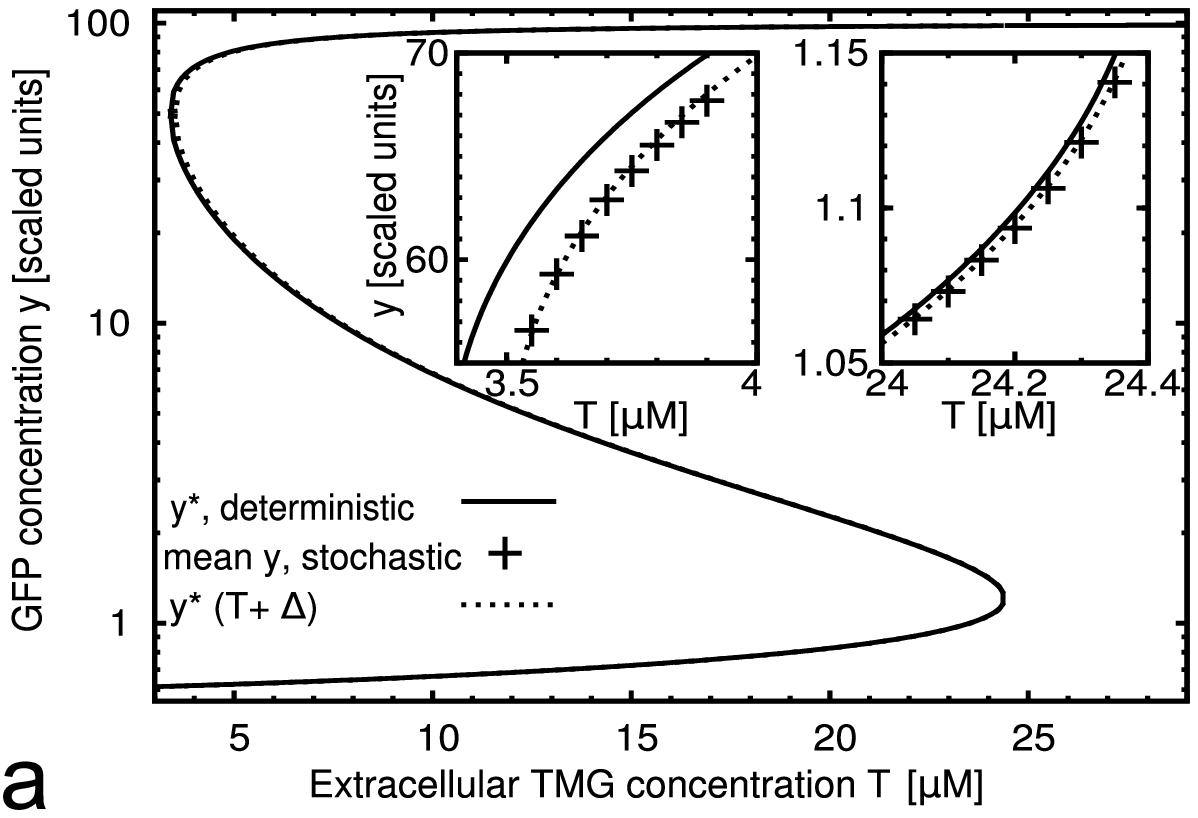} \hspace{0.5cm}
\includegraphics[width=7cm]{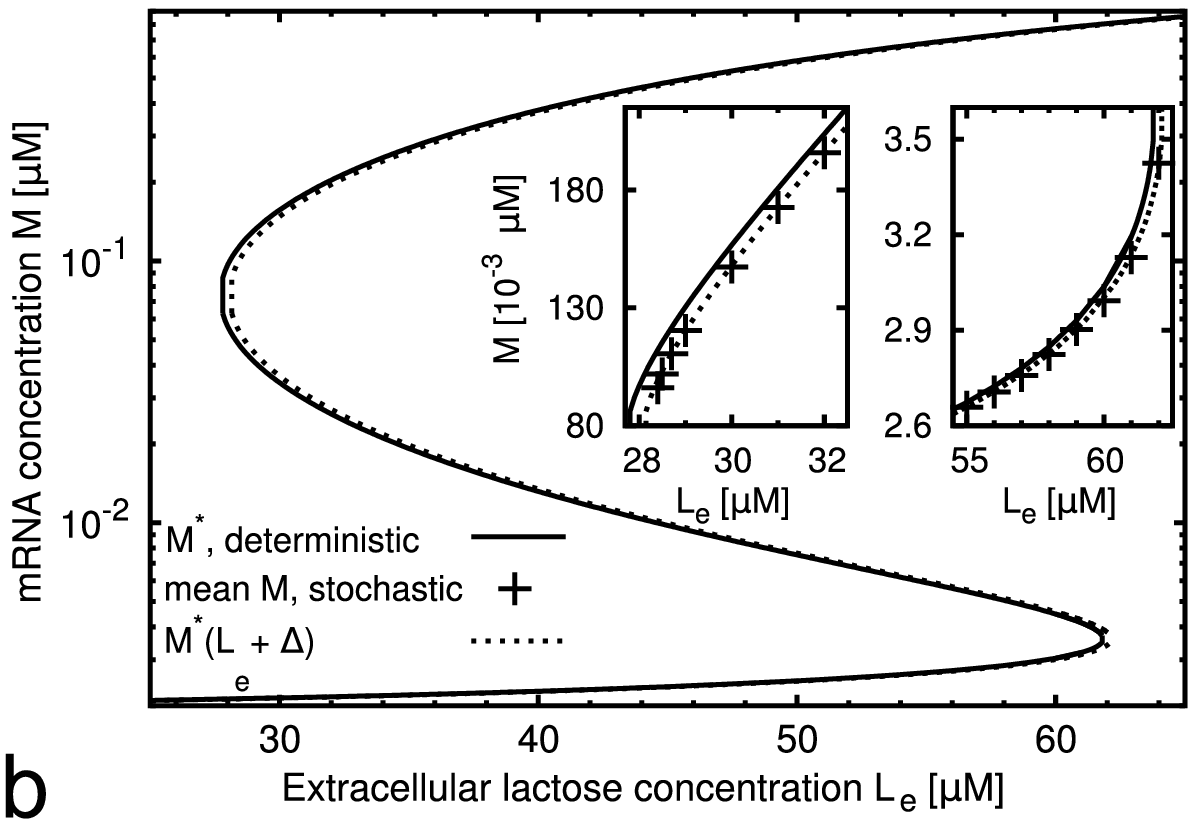}
\caption{\label{fig:shift} Comparison of the steady-state displacements (\ref{eq:shift_ozbudak}, \ref{eq:shift_mackey}) calculated using the small-noise expansion method with values obtained from the simulation of the Ozbudak model (a), and corresponding results for the Yildirim-Mackey model \citep{Ochab}  (b). The standard deviations of the fluctuations were $\sigma=1.5 \ \mathrm{\mu M}$ for the Ozbudak model and $\sigma=10 \ \mathrm{\mu M}$ for the Yildirim-Mackey model \citep{Yildirim_Mackey}. The response of both models to fluctuations in extracellular TMG/lactose concentration passing through a Michaelis-Menten-type uptake reaction is very similar: The steady states shift to the right. The displacement is larger for lower TMG/lactose concentrations and smaller for higher concentrations. }
\end{figure}

The steady-state displacement (\ref{eq:Delta}) does not depend on the kinetics of the system. It only depends on the input noise intensity and on the shape of the uptake function: its monotonicity and concavity. Therefore, only knowing the graph of the steady states vs. $\bar{P}$ , one can predict the changes in its stability due to noise passing through the uptake reaction.

For a monotonically increasing and concave uptake function, such as Michealis-Menten (Fig. \ref{fig:uptake}),  $h'>0$ and $h''<0$, so the steady states always shift to the right (in the direction of higher concentrations of P):
\be \label{eq:delta_right}
\Delta(\bar{P})  < 0
\ee
The formulas for TMG/lactose uptake rates (\ref{eq:tmg_uptake}), (\ref{eq:lactose_uptake}) are different in both models, but as they describe the same uptake mechanism, their graphs have a very similar Michaelis-Menten shape (Fig. \ref{fig:uptake}) with $h'>0$ and $h''<0$, which causes a steady-state shift to the right. In the Ozbudak model,
\be \label{eq:shift_ozbudak}
\Delta(\bar{T}) = - \frac{0.2}{\bar{T}} \sigma^2, 
\ee
while in the Yildirim-Mackey model \citep{Ochab},
\be \label{eq:shift_mackey}
\Delta(\bar{L}_e) = -\frac{1}{ (K_{L_e}+\bar{L}_e)}\sigma^2.
\ee

Within the bistable regions, the steady-state displacement due to noise in the Ozbudak model is smaller than in the Yildirim-Mackey model (Figs. \ref{fig:delta}, \ref{fig:shift}). A common feature of both models is a larger shift for low extracellular TMG/lactose concentrations and a smaller shift for high concentrations.

\begin{figure}[t]
\includegraphics[width=7cm]{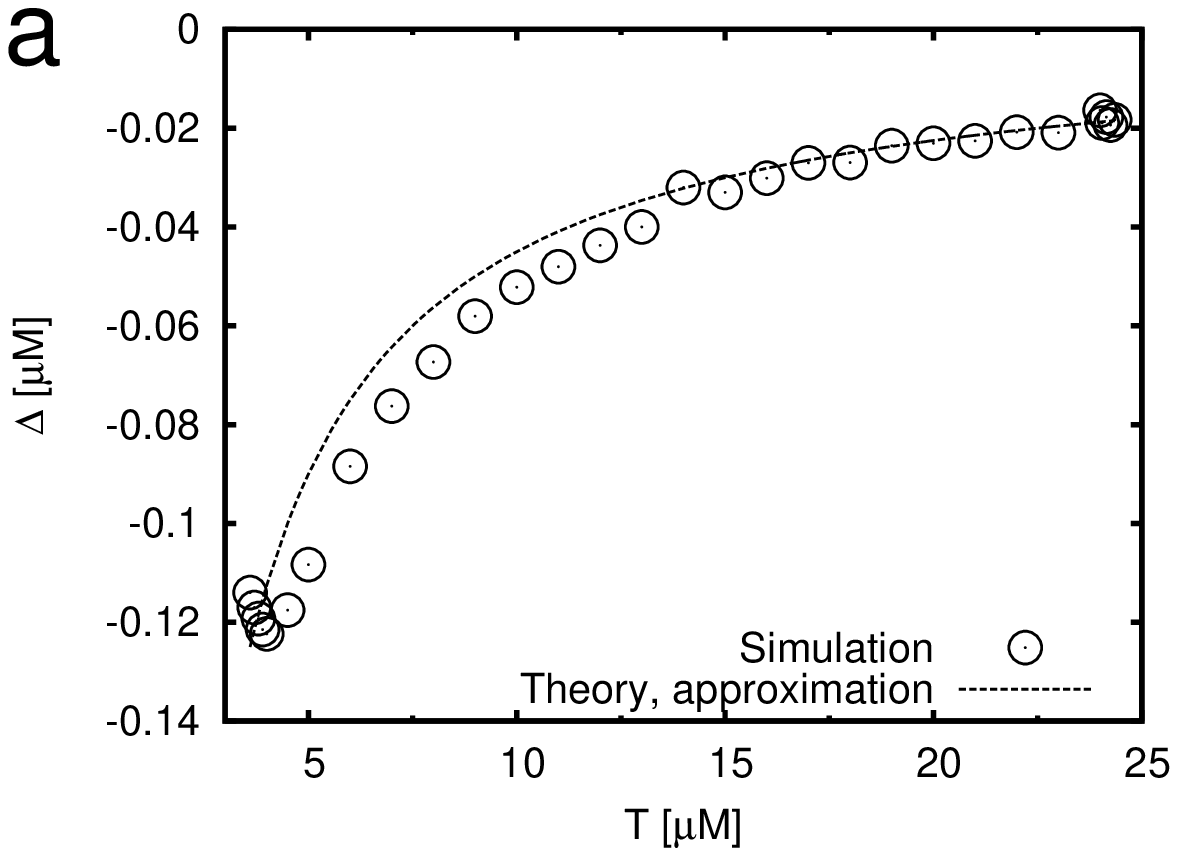} \hspace{0.5cm}
\includegraphics[width=7cm]{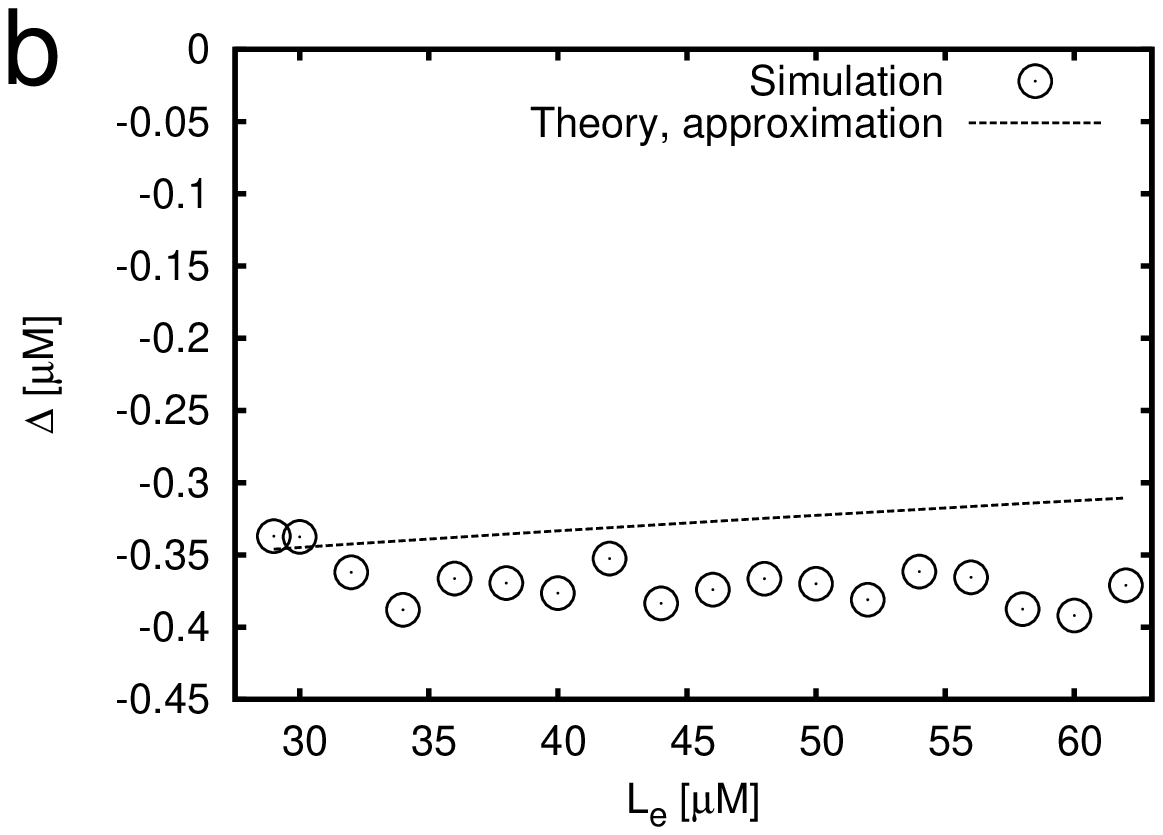}
\caption{\label{fig:delta_comp} The values of the steady-state displacement, calculated using the small-noise expansion method, compared with the values obtained from simulations. a) the Ozbudak model, b) the Yildirim-Mackey model. The noise intensities were same as in Fig. \ref{fig:shift}.}
\end{figure}

In the Ozbudak model as well as in the Yildirim-Mackey model, the results obtained analytically by small-noise expansion (\ref{eq:shift_ozbudak}, \ref{eq:shift_mackey}) were in a very good agreement with the simulation results (Figs. \ref{fig:shift}, \ref{fig:delta_comp}). The behavior of the Ozbudak model was  very similar to the behavior of the Yildirim-Mackey model: In both cases the graph of the steady states vs. the mean extracellular TMG/lactose concentration shifted to the right. The shift was larger for lower TMG/lactose concentrations and smaller for higher concentrations.

\subsection{Range of validity of the small-noise expansion}
For both models (within the given choice of parameters), we compared the values of $\Delta$ calculated using different methods for chosen constant values $\bar{T}$ or $\bar{L}_e$ as noise intensity and time scale of the noise were changed  (Fig. \ref{fig:delta_validity}). The values of $\Delta$ were: a) obtained from the simulation, b) calculated accurately (the mean $\langle h(P_t) \rangle$, computed using the distributions (\ref{eq:ph}) or (\ref{eq:pho}), was substituted into the equations of kinetics instead of $h(P_t)$), and c) calculated using the small-noise expansion. The results are consistent with those expected: The expansion (\ref{eq:expansion_Delta}) is valid when $\Delta \ll \bar{P}$, and indeed, in both models the approximation breaks down when $\Delta/\bar{T}$ or $\Delta / \bar{L}_e$ are greater than the order of $10^{-2}$. Moreover, the time scale of noise should be less than the time scale of the system, and for the Yildirim-Mackey model  the approximation is valid for $\tau_{OU}<0.1 \ \mathrm{min}$ while the fastest characteristic time for the left bifurcation point was $\tau_{sys} = 0.4 \  \mathrm{min}$ \citep{Ochab}. For the Ozbudak model, the approximation breaks down at  $\tau_{OU} \approx 0.1 \mathrm{min}$, while $\tau_{sys}=1 \mathrm{min}$. The results of the accurate calculation differ slightly from the small-noise approximation because of the Taylor expansion cut-off. There is also a systematic difference, increasing with noise intensity,  between the simulation results and those calculated accurately. This difference is due to the reflecting boundary conditions used in the simulations, which add a contribution from the trajectories reflected at $P_t=0$.

\begin{figure}[t]
\includegraphics[width=6.5cm]{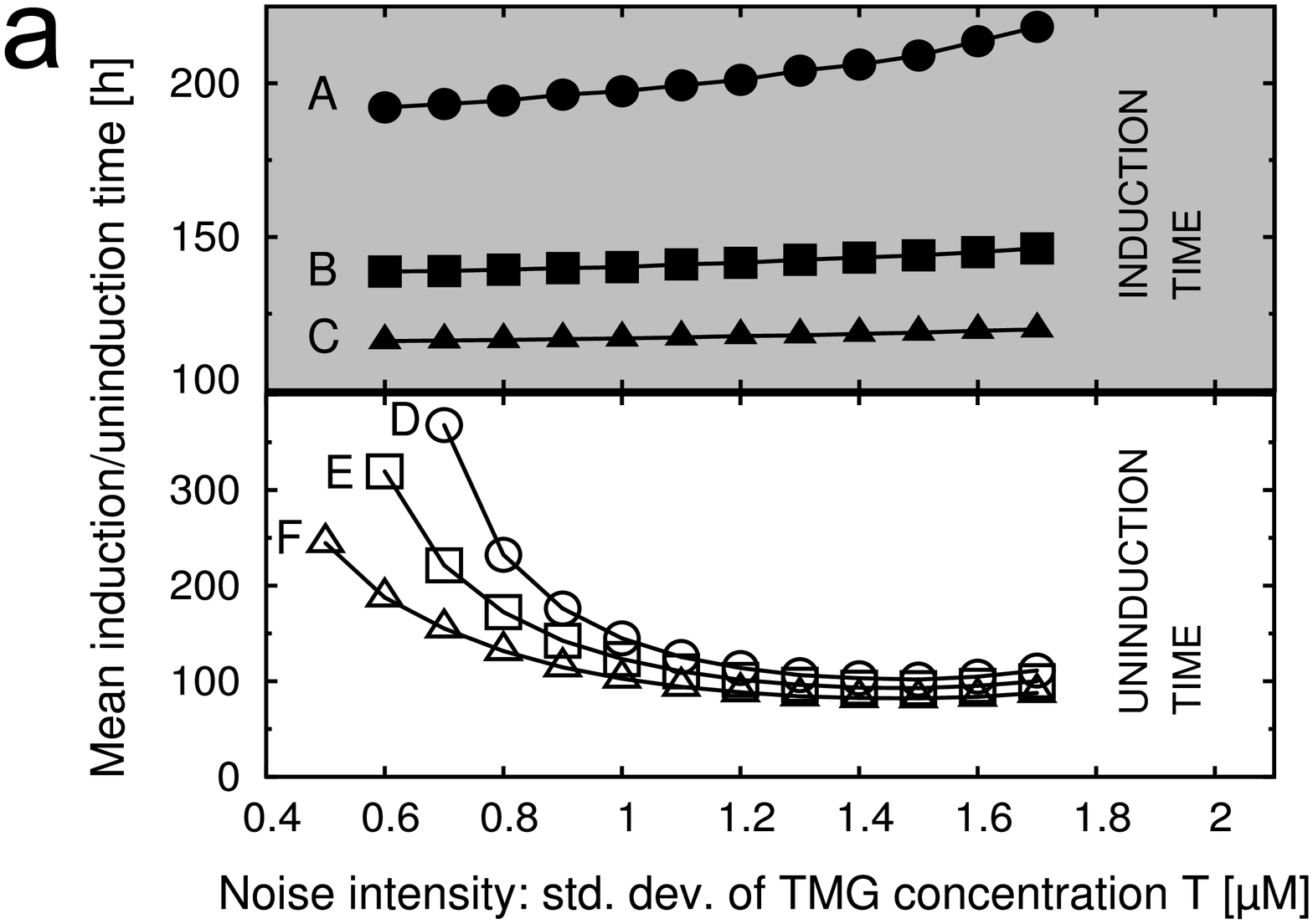}\hspace{0.5cm}
\includegraphics[width=7cm]{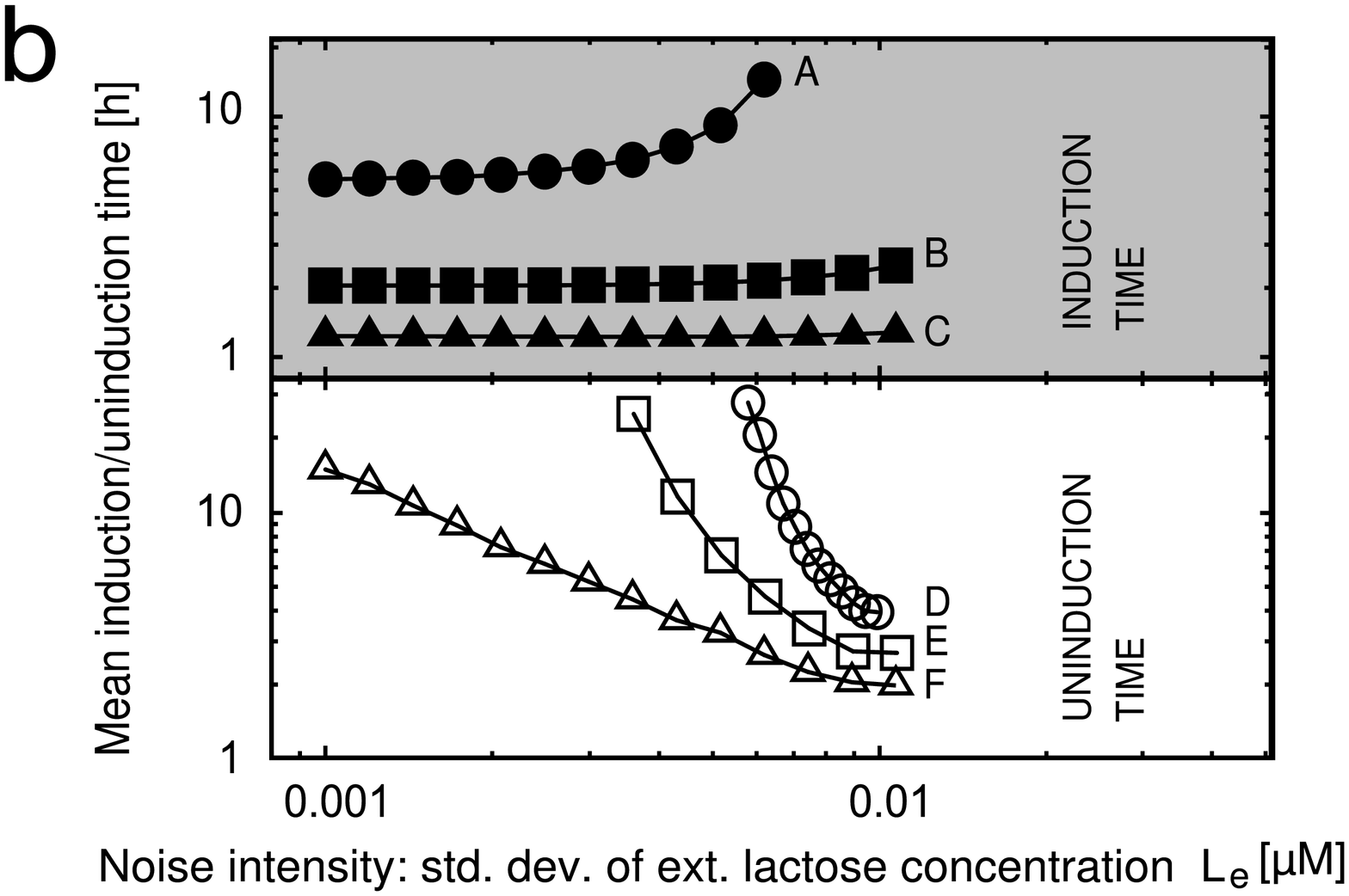}
\caption{\label{fig:mfpt} 
Increasing noise intensity inhibits induction and accelerates uninduction in the studied models of the \textit{lac} operon. Mean induction/uninduction time was measured in simulations of the Ozbudak model (a) with different noise intensities and different mean TMG concentrations $\bar{T}$: $24.5 \ \mathrm{\mu M}$ (A), $24.6 \ \mathrm{\mu M}$ (B), $24.7 \ \mathrm{\mu M}$ (C), $3.41 \ \mathrm{\mu M}$ (D), $3.40 \ \mathrm{\mu M}$ (E), $3.39 \ \mathrm{\mu M}$ (F). These results are compared with the results for the Yildirim-Mackey model (b) \citep{Ochab}, where the mean extracellular lactose concentrations $\bar{L}_e$ were: $62 \ \mathrm{\mu M}$ (A), $63 \ \mathrm{\mu M}$ (B), $65 \ \mathrm{\mu M}$ (C), $27.9 \ \mathrm{\mu M}$ (D), $27.8 \ \mathrm{\mu M}$ (E), $27.7 \ \mathrm{\mu M}$ (F). }
\end{figure}

\begin{figure}[t]
\includegraphics[width=7cm]{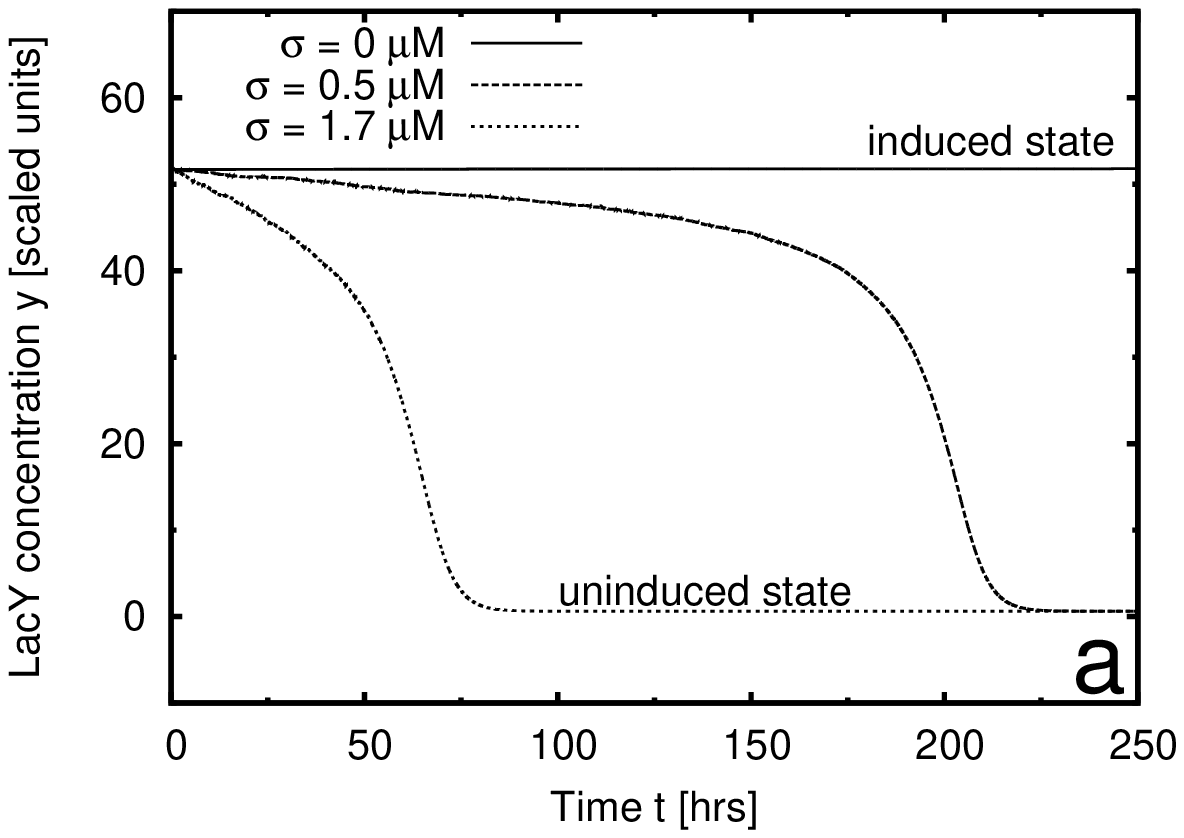}\\  
\includegraphics[width=7cm]{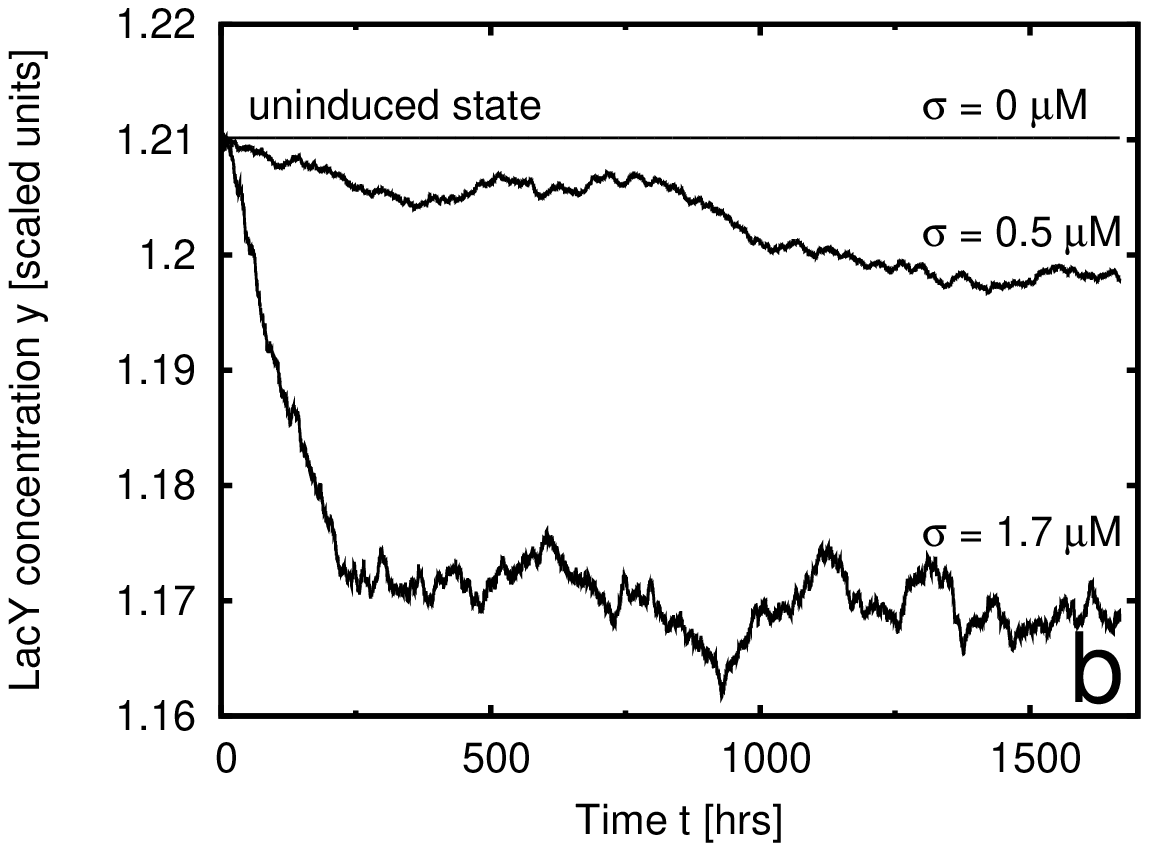} \includegraphics[width=7cm]{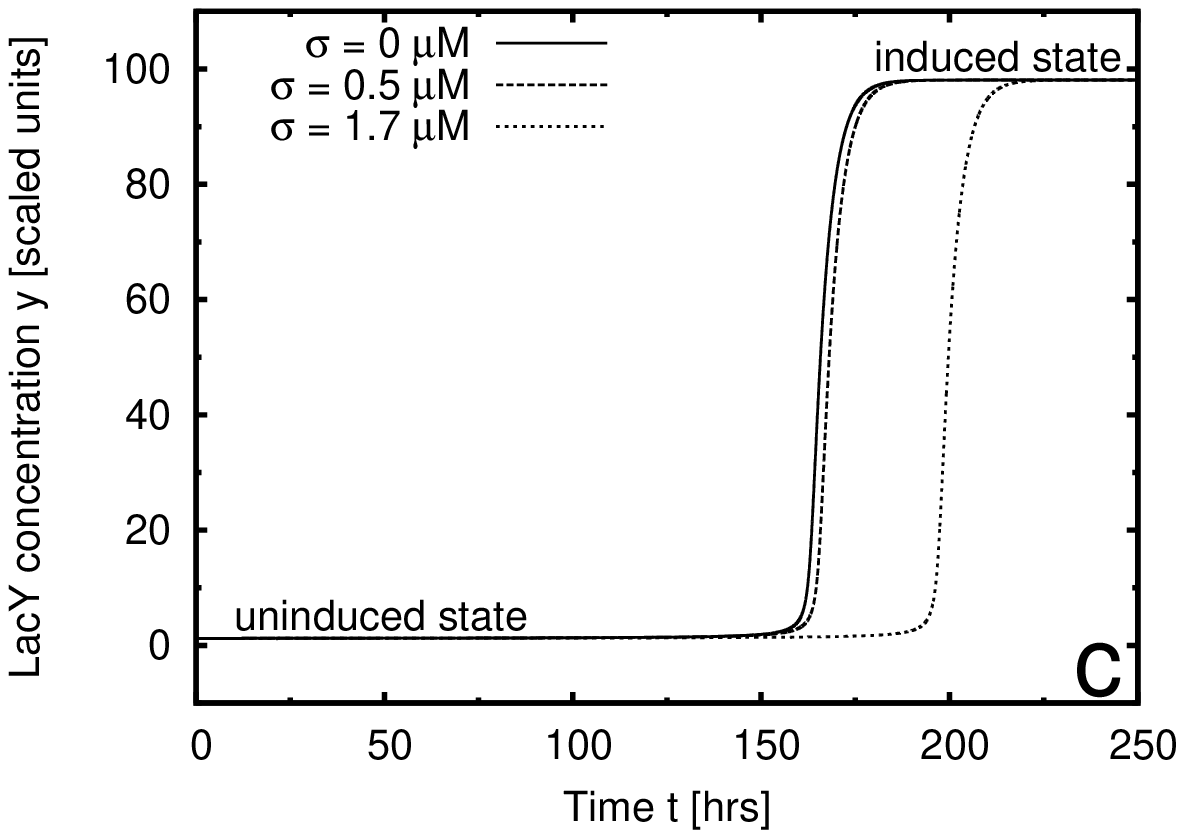}
\caption{\label{fig:traj} Increasing noise intensity inhibits induction and accelerates uninduction in the Ozbudak model of the \textit{lac} operon: Example trajectories for different initial states of the system: a) Induced, left bifurcation point ($\bar{T} = 3.39 \  \mu$M). b) Uninduced, right bifurcation point ($\bar{T} = 24.4 \ \mu$M). c) Uninduced, initial concentrations same as the coordinates of the right bifurcation point, but the TMG concentration is slightly beyond the bistability region ($\bar{T} = 24.5 \  \mu$M).}
\end{figure}

\subsection{Induction/uninduction time}

We compared the mean induction/uninduction times for both models, obtained from the simulations with different noise intensities and different TMG concentrations (Fig. \ref{fig:mfpt}). Both models are robust to fluctuations in extracellular TMG/lactose concentration. Noise-induced switching from the induced to uninduced state due to noise driving the system out of the steady state \citep{Horsthemke} was possible only for concentrations close to the boundaries of the bistable region. 

Therefore, to study the uninduction we performed simulations in which the system started within the bistable region, in the induced state very close to the left bifurcation point (Tab. \ref{tab:ozbudak_points}). When the noise intensity is zero, the system remains in the steady state and the uninduction time is infinite. But as the noise intensity increases, the bistable region shifts in the direction of larger TMG concentrations, effectively destabilizing the induced state and thus decreasing the mean uninduction time (Fig.~\ref{fig:traj}a).

On the other hand, within the bistable region the noise-driven switching from the uninduced steady state to the induced one was impossible in the range of the noise intensities for which the small-noise expansion  method is valid (Fig.~\ref{fig:traj}b). To observe the transitions to the induced state, we had to set the initial conditions outside the bistable region (Tab. \ref{tab:ozbudak_points}). The unstable initial positions become closer to the bistable region, and thus become more stable, as this region shifts in the direction of larger TMG concentrations. The trajectories spend more time in the vicinity of the starting point before they switch to the induced state, so the mean induction time increases as noise intensity increases (Fig.~\ref{fig:traj}c). This effect is same as in the Yildirim-Mackey model (Fig. \ref{fig:mfpt} b, \cite{Ochab}).

\section{Discussion}
In spite of the differences between the kinetics of the Ozbudak and Yildirim-Mackey models, and differently defined TMG/lactose uptake rate functions (the noise filters), the behavior of both models is qualitatively the same. The steady-state shift due to small noise and the consequent changes in the induction/uninduction times depend on the characteristics (monotonicity, concavity) of the filtering function and not on the details of the kinetics of the system.

Gaussian fluctuations of the extrinsic input signal, which enter the system through the Michaelis-Menten uptake reaction, generate an asymmetric distribution of its rate. This effect is similar as in \cite{Shahrezaei_Ollivier} and \cite{Lei}, where Gaussian noise enters the system through an exponential function. (Note that, however, in \cite{Shahrezaei_Ollivier} the noise was too slow (slower than the fastest time-scale of the system) and in \cite{Lei} the noise was too strong for the small noise-expansion to be valid.) While in the latter cases the rate distribution is log-normal, the distribution the Michaelis-Menten uptake rate has an opposite skewness. Due to the asymmetry of the distribution, the mean uptake rate varies as the input noise intensity is varied. This gives rise to the shift of the steady-state concentrations of the studied reaction system. 

The small-noise expansion turns out to be a useful tool for prediction of the steady-state displacement due to weak and rapid noise. Within the validity range of the method, the distribution of the input noise is not important (it can be non-Gaussian as well): It is only the shape of the uptake function that determines the direction and size of the steady-state displacement for a given input noise intensity. The simplicity of this approximate calculation allows for qualitative predictions, based on experimental data only, of the response of the studied system to extrinsic noise passing the uptake reaction of a given type. Assuming that we can only measure the intensity of the input noise, all information needed to predict the steady-state displacement are: a) Experimental the graph of the system's steady states vs. the mean input signal and b) Experimental graph of the uptake rate function vs. the mean input signal. Even if this data is not precise or already includes noise (i.e. is shifted due to noise), it enables qualitative predictions of the steady-state displacement  when noise intensity increases or decreases. The shape (monotonicity and concavity) of the uptake rate function indicates the direction of the noise-induced shift of the stationary states, whose approximate positions are known from the experimental graph. In particular, when the uptake rate is of Michaelis-Menten type (monotonically increasing and concave), the graph of the system's steady states vs. mean input signal always shifts to the right. In case of a bistable genetic switch, one can then predict that, as a consequence of such a shift, the extrinsic noise can selectively facilitate or inhibit induction and uninduction. 

In principle, it would be possible to measure the shift experimentally on the level of the Michaelis-Menten reaction rate $h(P) = P / (K_M + P)$. The shift could be detected as the noise correction (\ref{eq:Knoise}) to the Michaelis constant $K_M$, which could be read out from the experimental plot of $h(P)$. However, in the studied models of the \textit{lac} operon the noise would be probably too small for experimental detection of the shift of the reaction rate plot. But the computer simulations suggest that it would be easier to experimentally detect the shift by measurement of the switching times of the \textit{lac} switch.

The measurements of induction/uninduction times in the simulations of two example \textit{lac} operon models demonstrate that the bistability of the lactose utilization mechanism is robust to fluctuations in extracellular TMG/lactose concentration. However, even as small a noise as used in this study (to fulfill the conditions of validity of the small-noise expansion) can have a significant effect on the induction/uninduction time. The values by which it increases or decreases depend on the choice of model parameters and initial conditions. The effect is the strongest when the initial conditions are close to the bifurcation points (boundaries of the bistable region). For example in the Ozbudak model, when the initial state of the system is the left bifurcation point (induced state), then the uninduction time changes from infinity (at zero noise) to $\sim$100 hours (when the standard deviation $\sigma$ of the TMG fluctuations is 1.5 $\mu$M). On the other hand, when the initial state is the right bifurcation point (uninduced state), then the noise-driven induction is impossible. When the initial state is close to the right bifurcation point, but out of the bistable region, the induction time is increased by noise. For example, when the mean TMG concentration is 24.5 $\mu$M, then at zero noise the induction time is $\sim$190 hours, but at the noise of $\sigma = 1.7 \mu$M the induction time is $\sim$250 hours. These effects are qualitatively same as in the Yildirim-Mackey model \citep{Ochab}.

Thus, the fluctuations in extracellular TMG/lactose concentration facilitate the switching off of the TMG/lactose metabolism, but at the same time they prevent the metabolism from switching on. This effect will be valid for any model of \textit{lac} operon, provided that the TMG/lactose uptake rate is of Michaelis-Menten type. This suggests that in the presence of noise the possibility of random switching on the metabolism of TMG/lactose is more strongly protected than the possibility of random switching it off. One can speculate whether this protection against accidental induction of metabolism is connected with preventing an unnecessary energetic effort. To answer this question, one should analyze the lactose utilization system in \textit{E. coli} from the energetic point of view.

In other systems, a different direction of the noise-induced steady state displacement is possible. For example, when  extrinsic noise enters the system through the exponential function (monotinically increasing and convex), as in \cite{Shahrezaei_Ollivier} and \cite{Lei}, then one can predict that for weak and rapid noise the graph of the steady states vs. mean noise intensity will shift in the opposite direction than that of the systems with the Michaelis-Menten uptake. On the other hand, if the uptake rate is a Hill function $P^n / (K+P^n)$ then the formula (\ref{eq:Delta}) for the steady-state displacement can change its sign for different concentrations of $P$ and, under certain conditions, bistability may emerge due to noise in place of a graded response. Similar effects are possible for systems with non-monotonic input functions generated by incoherent feed-forward loops \citep{Kaplan, Kim}.

The analysis presented applies to weak and rapid noise from one dominating external source. But even in the systems where the contribution of other noises is present, the method may be of use to interpret the experimental measurements in terms of the discrimination between the effects of noises which originate from different sources.

\section{Acknowledgements}
The project was operated within the Foundation for Polish Science TEAM Program co-financed by the EU European Regional Development Fund: TEAM/2008-2/2. We would like to thank Dr. Paweł F. Góra (Jagiellonian University, Kraków, Poland) for valuable discussions and Edward Davis (The University of York, UK) for the help in improving the language of this article.

\newpage
\appendix
\section{Transformation of probability density functions}\label{app:pdf}
When $h(P)$ is an either monotonically increasing or decreasing function of a random variable $P$, and $P$ is given by the probability density function $q(P)$, then the probability density function of $h$ is given by the formula:
\be
p(h) = q(P(h)) \left| \frac{dP(h)}{dh} \right|,
\ee
where $P(h)$ is the inverse function of $h(P)$. The formula is obtained by the change of variables in the integration (see e.g. \cite{Miller}):
\begin{eqnarray}
\mathrm{Prob}(h(P_1) < h < h(P_2))  = \mathrm{Prob} (P_1 < P < P_2)  =\\ \nonumber 
 = \int_{P_1}^{P_2} q(P) dP  =  \int_{h(P_1)}^{h(P_2)} q(P(h)) \left| \frac{dP(h)}{dh} \right| dh =  \int_{h(P_1)}^{h(P_2)} p(h) dh 
\end{eqnarray}
(The absolute value is needed when $h(P)$ is decreasing.)

\section{Simulation details}\label{app:simul}
The fluctuations in the extracellular TMG/lactose concentration are modelled by the Ornstein-Uhlenbeck (OU) process \citep{Gardiner} with reflecting boundary conditions in $P_t=0$. $\xi(t)$ is a Gaussian white noise of intensity $\gamma$ and autocorrelation $\langle \xi(t) \xi(s) \rangle = \delta(t-s)$:
\be \label{eq:OU}
\frac{d P_t}{dt} = - \theta (P_t - \bar{P}) + \gamma \xi(t), \ \ P_t \geq 0. 
\ee
The correlation time of the noise $\tau_{OU} = 1 / \theta $. The variance of the OU process is $\sigma^2 = \gamma^2 / 2 \theta$.  We assume a small noise intensity and $\bar{P}$ sufficiently far from the reflecting boundary, so that the contribution of reflected 'tail' of the Gaussian distribution can be neglected and we can use formulas for the unbounded OU noise for the mean and variance. The noise intensity is varied in the simulations by varying the value of $\gamma$.  . Numerical integration of the equations of kinetics has been done using the Euler scheme \citep{nr,Mannella}. The timestep $\delta t=2\cdot 10^{-3} \ \mathrm{min}$ has been chosen significantly smaller than the time scales of the studied systems. 

 To estimate the mean induction/uninduction time in the simulations, the time was measured until trajectories $\vec{X}(t)$ got into a close neighborhood of the other deterministic stationary state (of a radius $D = \sqrt{\sum_i \delta X_i^2} = 0.1$ [scaled  units] for the Ozbudak model and $D=5 \ \mathrm{\mu M}$ for the Yildirim-Mackey model \citep{Ochab}). The initial points of the trajectories $\vec{X}$ were the deterministic steady states within the bistable region, or points outside the bistable region, but close to its boundaries (see Tables \ref{tab:ozbudak_points}, \ref{tab:mackey_points}). The number of simulation runs for calculating the mean induction/uninduction time was $N=100$ for the Ozbudak model and $N=1000$ for the Yildirim-Mackey \citep{Yildirim_Mackey} model.


\section{Tables}


\begin{table}[!h]
\begin{tabular}{|l|l|} \hline
$\rho=1+\frac{R_T}{R_0}$ & 167.1  $^a$\\
$\alpha$                 & 100.5  $^a$\\
$\beta_G (G=0)$                  & 100 $^a$\\
$\tau_y$                 & 216 min $^b$\\
$\tau_x$                 & 1 min $^c$\\
\hline
\end{tabular}
\caption{\label{tab:ozbudak_params}
Parameters of the Ozbudak model: $^a$) \citep{Ozbudak}, $^b$) \citep{Mettetal},
$^c$) chosen for this study within the range of $0 .. 35$ min reported by \citep{Mettetal}.
} 
\end{table}

\begin{table}[!h]
\begin{tabular}{|l|l||l|l|} \hline
$\Gamma_0$ & $7.25 \times 10^{-7} \ \mathrm{mM/min}$&$\mu$ & $0.0226 \ \mathrm{min}^{-1}$\\
$\alpha_A$ & $1.76 \times 10^4 \ \mathrm{min}^{-1}$ &$\tau_B$ & $2.0 \ \mathrm{min}$\\
$\alpha_B$ & $1.66 \times 10^{-2} \ \mathrm{min}^{-1}$ &$\tau_M$ & $0.1 \ \mathrm{min}$\\
$\alpha_L$ & $2.88 \times 10^3 \ \mathrm{min}^{-1}$ & $\tau_P$& $0.83 \ \mathrm{min}$\\
$\alpha_M$ & $9.97 \times 10^{-4} \ \mathrm{mM/min}$ & $K$& $7.2 \times 10^3$\\
$\alpha_P$ & $10.0 \ \mathrm{min}^{-1}$& $K_1$& $2.52 \times 10^4 \ \mathrm{mM}^{-2}$\\
$\beta_A$ & $2.15 \times 10^4 \ \mathrm{min}^-1$ & $K_A$ & $1.95 \ \mathrm{mM}$\\
$\beta_L$ & $2.65 \times 10^3 \ \mathrm{min}^{-1}$ & $K_L$& $0.97 \ \mathrm{mM}$\\
$\gamma_A$ & $0.52 \ \mathrm{min}^{-1}$ & $K_{L_e}$& $0.26 \ \mathrm{mM}$\\
$\gamma_B$ & $8.33 \times 10^{-4} \ \mathrm{min}^{-1}$ & $K_{L_1}$& $1.81 \ \mathrm{mM}$ \\
$\gamma_L$ & $0.0 \ \mathrm{min}^{-1}$ & $k_B$& $0.677$\\
$\gamma_M$ & $0.411 \ \mathrm{min}^{-1}$& $k_P$& $13.94$\\
$\gamma_P$ & $0.65 \ \mathrm{min}^{-1}$& &\\
\hline
\end{tabular}
\caption{\label{tab:mackey_params} Parameters of the Yildirim-Mackey model \citep{Yildirim_Mackey}.}
\end{table}

\begin{table}[!h]
\begin{tabular}{|l|l|l|l|l|}
\hline
$\bar{T} [\mathrm{\mu M}]$& $x_i [\mathrm{scaled \ units}]$ & 
$y_i [\mathrm{scaled \ units}]$ & $x_f [\mathrm{scaled \ units}]$ & 
$y_f [\mathrm{scaled \ units}]$ \\
\hline
A: 24.5 $^a$ & 1.012 & 1.210 & 82.23 & 98.09\\
B: 24.6 $^a$ & 1.012 & 1.210 & 82.44 & 98.10\\
C: 24.7 $^a$ & 1.012 & 1.210 & 82.65 & 98.12\\
D: 3.39 &13.228  & 51.700 &0.1577  & 0.6163\\
E: 3.40 & 13.707 & 53.475 &  0.1580 & 0.6164 \\
F: 3.41 & 14.012 & 54.569 &  0.1583&  0.6164\\
\hline
\end{tabular}
\caption{\label{tab:ozbudak_points} Initial ($i$ subscript) and final ($f$ subscript) concentrations of intracellular TMG ($x$) and permease ($y$) for the simulation measurements of mean switching time in the Ozbudak model. Trajectories A, B, C started from outside the bistable region, but very close to its boundaries. Initial points for these trajectories are the coordinates of the right ($^a$) bifurcation point.)}
\end{table}

\begin{table}[!h]
\begin{tabular}{|l|l|l|l|l|l|l|l} 
\hline
$\bar{L}_e [\mathrm{\mu M}]$& $A_i [\mathrm{\mu M}]$ & 
$L_i [\mathrm{\mu M}]$ & $M_i [10^{-2} \mathrm{\mu M}]$ &
$A_f [\mathrm{\mu M}]$ & $L_f [\mathrm{\mu M}]$ & $M_f [10^{-2} \mathrm{\mu M}]$ \\ 
\hline
$A: 62.0^a$ & $14.1$  &$224$ & $0.360$ & $393$ & $285$ & $80.8$\\
$B: 63.0^a$ & $14.1$  &$224$ & $0.360$ & $399$ & $289$ & $82.5$\\
$C: 65.0^a$ & $14.1$  &$224$ & $0.360$ & $412$ & $298$ & $85.9$\\
$D: 27.9$   & $109.0$ &$129$ & $9.406$ & $4.00$   & $94.6$  & $0.212$\\
$E: 27.8$   & $104.0$ &$129$ & $8.600$ & $4.00$   & $94.0$  & $0.212$\\
$F: 27.7$   & $96.9$  &$128$ & $7.521$ & $3.98$   & $93.7$  & $0.212$\\
\hline
\end{tabular}
\caption{\label{tab:mackey_points} Initial ($i$ subscript) and final ($f$ subscript) concentrations of allolactose ($A$), lactose ($L$), and mRNA ($M$) for the simulation measurements of mean switching time in the Yildirim-Mackey model \citep{Yildirim_Mackey}. $^a$) Trajectories starting from outside the bistable region, but very close to its boundaries. Initial points for these trajectories are the coordinates of the right bifurcation point.}
\end{table}

\bibliographystyle{elsart-harv}

\bibliography{praca09}

\end{document}